\documentclass{jpsj3}
\usepackage{txfonts}
\usepackage{color}
\usepackage{setspace}

\title{
Detection of phase transition in generalized P\'{o}lya urn 
\\ in information cascade experiment
}

\author{
Masafumi Hino$^1$, 
Yosuke Irie$^2$, Masato Hisakado$^3$, Taiki Takahashi$^4$, 
and Shintaro Mori$^5$\thanks{mori@sci.kitasato-u.ac.jp}
}

\inst{$^1$
NEC Corporation\\
Siba 5-7-1, Minato-ku, Tokyo 108-8001, Japan\\
$^2$Advanced Traffic Systems Corporation\\ 
Nishisinjuku 1-22-2, Shinjuku-ku, Tokyo 160-0023, Japan\\
$^3$Financial Services Agency\\
Kasumigaseki 3-2-1, Chiyoda-ku, Tokyo 100-8967, Japan\\
$^4$Department of Behavioral Science, Faculty of Letters
\\
Center for Experimental Research in Social Sciences, Hokkaido University
\\
Kita 10, Nishi 7, Kita-ku, Sapporo, Hokkaido 060-0810, Japan\\
$^5$Department of Physics, Kitasato University \\
Kitasato 1-15-1, Sagamihara,  Kanagawa 252-0373, Japan
} 

\abst{
 We propose a method of detecting a phase transition in
 a generalized P\'{o}lya urn in an information cascade experiment.
 The method is based on the asymptotic behavior of the
 correlation $C(t)$ between the first subject's choice and
 the $t+1$-th subject's choice, the
 limit value of which, $c\equiv \lim_{t\to \infty}C(t)$, is the order
 parameter of the phase transition.
 To verify the method, we perform a voting
 experiment using two-choice questions.
 An urn X is chosen at random from two urns A and B, which contain
 red and blue balls in different configurations.
 Subjects sequentially guess whether X is A or B using information about the
 prior subjects' choices and the color of a ball randomly drawn from X.
 The color tells the subject which is X
 with probability $q$. We set
$q\in \{5/9,6/9,7/9,8/9\}$ by controlling the
 configurations of red and blue balls in A and B.
 The (average) lengths of the sequence of the subjects
 are 63, 63, 54.0, and 60.5 for $q\in \{5/9,6/9,7/9,8/9\}$, respectively.
 We describe the sequential voting process by a
 nonlinear P\'{o}lya urn model. The model suggests the possibility of
 a phase transition when $q$ changes.
 We show that $c>0\,\,\,(=0)$ for $q=5/9,6/9\,\,\,(7/9,8/9 )$ and
 detect the phase transition using the proposed method.
}

\begin{document}
\maketitle

\section{\label{sec:intro}Introduction}
The social contagion process has long been extensively studied 
\cite{Wat:2007,Sum:2009,Gra:2014}.
Because of progress in information communication technology,
we often rely on social information for decision making
\cite{Sal:2006,Bon:2012,Wan:2014}.
The P\'{o}lya urn is a simple stochastic process in which contagion is
taken into account by a reinforcement mechanism \cite{Pol:1931}.
There are initially $R_{0}$ red balls and $B_{0}$ blue balls in an urn.
At each step, one draws a ball randomly from the urn and duplicates it.
Then, one returns the balls, and the probability of selecting
a ball of the same color
 is strengthened.
As the process is repeated infinitely,
the ratio of red balls
 in the urn $z$ becomes random and obeys
 the beta distribution $\beta(R_{0},B_{0})$.
 In the process, information on the first draw propagates and affects
 infinitely later draws. The correlation between the color of
 the first ball and
 that of a ball chosen later
 is $1/(R_{0}+B_{0}+1)$ \cite{His:2006}.

 As the P\'{o}lya urn process is very simple, and
 there are many reinforcement
 phenomena in nature and the social environment, many variants of the process
 have been proposed under the name of generalized P\'{o}lya urn
 \cite{Pem:2007}.
 One example is the lock-in phenomenon proposed by Arthur as
 a mechanism by which a technology, product, or service dominates others and
 occupies a large market share \cite{Art:1989}.
 The dominant one is not necessarily superior to the others in some respect.
 The necessary condition for lock-in
 is externality, in which wider adoption induces posterior superiority.
 Arthur used a generalized P\'{o}lya urn to explain the lock-in
 phenomenon. In the process, the choice of the ball (technology, product, or
 service) is described by a nonlinear
 function $f(z)$ of the ratio of red balls $z$.
 In contrast to the original P\'{o}lya
 urn, where $f(z)=z$, the ratio of red balls
 converges to a stable fixed point $z_{*}=f(z_{*})$ in the nonlinear model
 \cite{Hil:1980}.
 Mathematically, the fixed points $z_{*}$
 are categorized as upcrossings and downcrossings,
 at which the graph $y=f(z)$ crosses
 the graph $y=z$ going upward and downward, respectively.
 The downcrossing (upcrossing) fixed point is stable (unstable),
 as the probability that $z$ converges to it is positive (zero).
 Arthur adopted an S-shaped $f(z)$ with two stable fixed points and
 noted that random selection among the fixed points
 also occurs in the adoption process.

 If the number of stable fixed points changes as one changes
 the parameters of the function $f(z)$,
 the generalized P\'{o}lya urn shows a transition
 \cite{His:2011,His:2012}.
 The order parameter is the limit value of the
 correlation between the first drawn ball
 and later drawn balls \cite{Mor:2015,Mor:2015-2}.
 If $f(z)$ is $Z_{2}$-symmetric and satisfies
 $f(z)=1-f(1-z)$,
 the transition becomes continuous, and
 the order parameter satisfies a scaling relation
 in the nonequilibrium  phase transition.
 One good candidate for experimental
 realization of the phase transition
 is  the information cascade experiment \cite{Bik:1992}.
 There, participants answer two-choice questions
 sequentially. In the canonical setting of the experiment,
 two urns, A and B,
 with different configurations of red and blue balls are
 prepared \cite{And:1997,Kub:2004,Goe:2007}.
 One of the two urns is chosen at random to be urn X,
 and the question is whether urn X is A or B.
 The participants can draw a ball from urn X and
 see which type of ball it is.
 This knowledge, which is called the private signal,
 provides some information about X.
 However, the private signal does not indicate the true situation
 unequivocally, and participants have to decide under uncertainty.
 Participants are also provided with social information
 regarding how many prior participants
 have chosen each urn. The social information
 introduces an externality to the decision making: as
 more participants choose urn A (B), later 
 participants are more likely to identify urn X as urn A (B).
 The social interaction in which a participant tends to 
 choose the majority choice
 even if it contradicts the private signal is called an information
 cascade or rational herding \cite{Bik:1992}.
 In a simple model of information cascade,
 if the difference in the numbers of subjects who have chosen
 each urn  exceeds two, the social
 information overwhelms subjects' private signals.
 In the limit of many previous subjects, the decision
 is described by a threshold rule
 stating that a subject chooses an option
 if its ratio exceeds $1/2$, $f(z)=\theta(z-1/2)$.
 The function $f(z)$ that describes decisions under
 social information is called a response function
\cite{Wat:2002}.

 To detect the phase transition caused by the change in $f(z)$,
 we have proposed another information cascade experiment
 in which subjects answer two-choice general knowledge
 questions \cite{Mor:2012,Mor:2013}.
 If almost all of the subjects know the answer to a question, the probability
 of the correct choice is high, and $f(z)$ does not depend
 greatly on the social information.
 In this case, $f(z)$ has only one stable fixed point.
 However, when almost all the subjects do not know
 the answer, they show a strong tendency to choose the majority answer.
 Then $f(z)$ becomes S-shaped, and it could have multiple stable fixed points.
 We have shown that when the
 difficulty of the questions is changed,
 the number of stable fixed points of
 the experimentally derived $f(z)$ changes \cite{Mor:2012}.
 If the questions are easy, there is only one stable fixed point,
 $z_{+}$, and the ratio of the correct choice $z$ converges
 to that value. If the questions are difficult,
 two stable fixed points, $z_{+}$ and $z_{-}$, appear.
 The stable fixed point to which
 $z$ converges becomes random.
 To detect the randomness using experimental data,
 we study how the variance of $z$ changes
 as more subjects answer questions of fixed difficulty.
 We showed that the variance
 converges to zero in the limit of many subjects for easy questions.
 For difficult questions, it converges to a finite and
 positive value, which suggests the
 existence of multiple stable states in the system.

 In this paper, we propose a new method of detecting
 the phase transition of a nonlinear P\'{o}lya urn
 in an information cascade experiment. It is based on
 the asymptotic behavior of the correlation function and the estimation
 of its limit value.
 We perform an information cascade experiment to verify 
 our method.
 We adopt the canonical setting for an information cascade experiment,
 in which subjects
 guess whether urn X is urn A or urn B.
 In the proceedings of ECCS'14, we
 reported some results
 from the present experiment \cite{Mor:2015-3}.
 Here, we provide complete information about the proposed method and the
 results of analysis of the experimental data.

 The paper is organized as follows.
 Section \ref{sec:model} considers a simple
 model of information cascade. We estimate the correlation function and
 the order parameter.
 In Sect. \ref{sec:setup},
 we explain the experimental procedure. Section \ref{sec:da} presents
 the analysis of the experimental data. We propose a nonlinear P\'{o}lya
 urn model based on the empirically estimated response function
 in Sect. \ref{sec:da2}. We estimate the order parameter
 by extrapolating the experimental results to a 
 larger system. We show the possibility of the phase transition
 in the thermodynamic limit.
 Section \ref{sec:con} presents a summary and future problems.
 Appendices provide additional information about the experiments.

\section{\label{sec:model}Simple Model of Information Cascade}
 We study a simple model of information cascade, which is a modification 
 of the "Basic model" in \cite{Bik:1992}. Assume that there are two options,
 A and B, one of which is chosen to be correct with equal probability.
 Each individual privately observes a conditionally
 independent signal about the true option.
 Individual $i$'s signal, $S_{i}$, is A or B, and A is observed
 with probability
 $q$ if the true option is A and with probability $1-q$ if the true option
 is B. Each individual also observes the decisions of all those ahead of him.
 Without loss of generality, we label the correct (incorrect) 
 option as 1 (0), and
 $S_{i}\in \{0,1\}$. The probability that $S_{i}=1$ is $q$.

 We assume that the first individual chooses 1 
 (0) if his private signal is 1 (0).
 The second individual can infer the first individual's signal from
 his decision. If the first individual chose 1 (0), the second individual
 chooses 1 (0) if his signal is 1 (0). If his signal contradicts
 the first individual's choice, we assume he chooses the same option
 as his signal, which is different from the tie-breaking convention
 in the "Basic model" \cite{Bik:1992},
 where the individual chooses 1 or 0 with equal probability.
 There are three situations for the third individual:
 (1) Both predecessors have chosen 1. Then, irrespective of his signal,
 he chooses 1. The following individuals also choose 1 and
 a correct cascade, which is called an up cascade in \cite{Bik:1992},
 starts.
 (2) Both have chosen 0, and an incorrect cascade, or down cascade, starts.
 (3) One has chosen 1, and the other has chosen 0. 
 The third individual is in the same
 situation as the first individual, and he choose the option
 matching his signal. The probability that both of the first two individuals
 receive correct (incorrect) signals is $q^{2} ((1-q)^{2})$, so an up
 (down) cascade starts with probability $q^{2} ((1-q)^{2})$.

\begin{figure}[htbp]
\begin{tabular}{c}
\includegraphics[width=10cm]{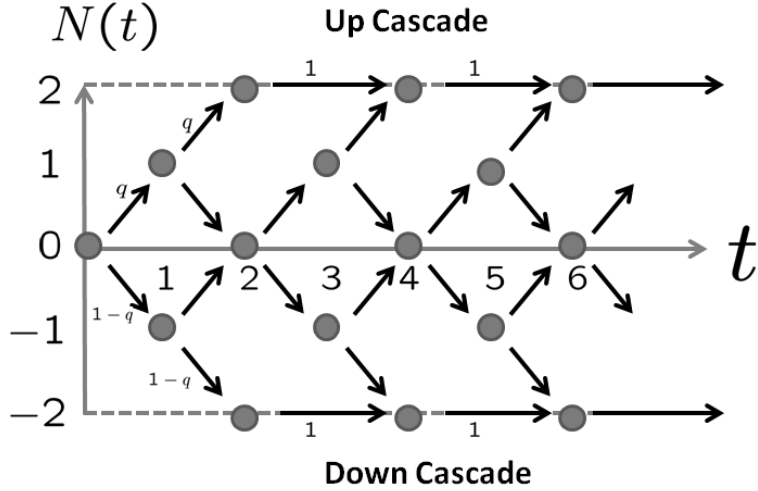}
\end{tabular}
\caption{\label{fig1} Simple model of information cascade.
 States $N(t)\in \{2,1,0,-1,-2\}$
 and probabilities for $X(t)\in \{0,1\}$.
}
\end{figure}	

 We denote the difference in the number of
 correct and incorrect choices up to the $t$-th individual as
 $N(t)$. From the above discussion, if $N(t)\ge 2 (\le -2)$,
 an up (down) cascade starts. There are essentially
 five states, $N(t)\in \{-2,-1,0,1,2\}$, if we identify
 all states with $N(t)\ge 2 (\le -2)$ as $N(t)=2 (-2)$.
 If $t$ is even, there are three states, $N(t)\in \{-2,0,2\}$, and
 there are four states, $N(t) \in \{-2,-1,1,2\}$, if $t$ is odd.
 Figure \ref{fig1} illustrates the model.
 In the figure, we also show the probabilistic rule for
 the transition between states.
 At $t=0$, $N(0)=0$, and it jumps to $N(t)=1 (-1)$
 with probability $q\,\, (1-q)$.
 From $t=1$ to $t=2$, the same rule applies, and $N(t)$ increases (decreases)
 by 1 with probability $q \,\,(1-q)$. If $N(t)=2 (-2)$ at $t=2$, an up (down)
 cascade starts. Later individuals choose 1 (0) for $t\ge 3$,
 and $N(t)$ remains $2 (-2)$.
 If $N(t)=0$ at $t=2$, the third individual chooses 1 with probability $q$.
 In general, if $|N(t)|\le 1$, $N(t)$ increases (decreases) by
 1 with probability $q\,\, (1-q)$.
 The problem is a random walk model with absorbing walls at $N(t)=\pm 2$.
 As $t$ increases, the probability that the random walk is absorbed in the
 walls increases. In the limit $t\to \infty$, all random walks are absorbed
 in the walls. The state $N(t)=0$ for even $t$ is
 absorbed into the state $N(t+2)=2$
 with probability  $q^{2}/(q^{2}+(1-q)^{2})$ 
 and is absorbed into the state  $N(t+2)=-2$ with probability
 $(1-q)^{2}/(q^{2}+(1-q)^{2})$.
 The probability for an up cascade in the
 limit $t\to \infty$, which we denote by $P_{2}(\infty)$,
 is then given as
\begin{equation}
P_{2}(\infty)\equiv
\mbox{Pr}(N(\infty)=2)=\frac{q^{2}}{q^{2}+(1-q^{2}} \label{eq:P2}.
\end{equation}
In the up (down) cascade, individuals always choose 1 (0), and
$P_{2}(\infty)$ is the limit value for the
probability of the correct choice.
It is greater than $q$ for $q>1/2$, and the deviation
shows an increase in the accuracy from
that  of the signal.
$P_{2}(\infty)-q$ is a measure of the collective intelligence.

We denote the $t$-th individual's choice as $X(t)\in \{0,1\}$.
We are interested in the estimation of the correlation function $C(t)$,
which is defined as the covariance of $X(1)$ and $X(t+1)$ divided by
the variance of $X(1)$.  $C(t)$ can also be defined as the difference
 in the conditional probabilities:
\[
C(t)=\mbox{Pr}(X(t+1)=1|X(1)=1)-\mbox{Pr}(X(t+1)=1|X(1)=0).
\]
$C(t)$ is then estimated as
\begin{eqnarray}
C(2n)&=&c(q)+
\frac{(1-2q)^{2}}{2(q^{2}+(1-q)^{2})}(\sqrt{(2f(1-q))})^{2n}, \nonumber \\
C(2n+1)&=&C(2n), \nonumber \\
c(q)&=&\lim_{t\to \infty}C(t)=\frac{q(1-q)}{q^{2}+(1-q)^{2}}.  \label{eq:c}
\end{eqnarray}
The derivation of $C(t)$ is given in appendix \ref{sec:app:model}.
The limit value $c(q)=\lim_{t\to\infty}C(t)$ is
the order parameter of the phase transition in
a nonlinear P\'{o}lya urn.
The order parameter $c(q)$ changes continuously with $q$, and
it takes zero at $q=0,1$.
The simple model does not show a phase transition,
 and $C(t)$ decays exponentially with $t$.

\section{\label{sec:setup}Experimental Setup}
The experiments reported here were conducted at Kitasato University.
We performed two experiments, EXP-I and EXP-II.
In EXP-I (II), we recruited $|ID|$ = 307 (33) students, mainly
from the School of Science.
In EXP-I (II), we prepared $I=200 (33)$ questions for
$q\in Q=\{5/9,6/9,7/9\} (8/15,5/9,6/9)$ and $I=400$ questions
for $q=8/9$. 
EXP-I was performed during three periods, $q\in \{5/9,6/9\}$
in  2013, $q=7/9$ in 2014, and $q=8/9$ in 2015.
EXP-II was performed in  2011.
We label the questions as $i=1,2,\cdots,I$.
 Subjects answered $I/2 \, (I)$ questions for some
(all) values of  $q$ in $Q$ in EXP-I (II).
We obtained $I$ sequences of answers
of length $T=63 \,(33)$ for $q=5/9,6/9 (8/15,5/9,6/9)$ in EXP-I (II).
In EXP-I for $q=7/9$ and $q=8/9$,
some subjects could not answer
$I/2$ questions within the allotted time.
The length $T$ of the sequence
depends on $i$, and the average (minimum)
length $T_{avg} (T_{min})$ is 54.0 (49) for $q=7/9$ and
60.5 (58) for $q=8/9$.

 $|ID|$ subjects sequentially answered a two-choice question and
 received returns for each correct choice.
 We prepared $I$ questions for each $q\in Q$ by randomly choosing an urn
 from two different urns, urn A and urn B, which contain ball A
 (red) and ball B (blue) in different proportions.
 We denote the answer to question $q\in Q,i \in \{1,\cdots,I\}$ as
 $U(q,i)\in \{A,B\}$.
 For $q=n/m>1/2$, urn A (B) contains $n$ A (B) balls
 and $m-n$ B (A) balls. Urn A (B) contains more
 A (B) balls than B (A) balls.
 The subjects obtain information about urn X by
 knowing the color of a ball randomly drawn from it.
 The color of the ball is the private signal, as it is not shared
 with other subjects.
 If the ball is ball A (B), X is more likely to
 be A (B). Further, $q$ is the posterior probability
 that the randomly chosen ball
 suggests the correct urn and the private signal is correct.
 We prepared the private signal $S(q,i,t)\in \{A,B\}$
 for $T$ subjects and $I$ questions in advance.
 In EXP-I, we controlled the ratio of the correct signal
 so that it was precisely $q$. Among $T$ subjects,
 exactly $q\cdot T$ subjects
 received the correct signal.
 In EXP-II, we did not control the private signal.
 Among 33 subjects, $q\cdot 33$ subjects
 received the correct signal on average.
 Table \ref{tab:design} summarizes the design.

\begin{table*}[htbp]
\caption{\label{tab:design}
Experimental design. $|ID|$, number of subjects; $T$, length of
private signal; $T_{avg}$,
average length of subject sequence; $T_{min}$, minimum length of
subject sequence; $\{q\}$,
precision of private signal; $I$, number of questions.}
\begin{tabular}{lcccccc}
\hline
Experiment & $|ID|$ & $T$ & $T_{avg}$ & $T_{min}$ & $\{q\}$ & $I$   \\
\hline
I (2013.9$\sim$ 2013.10) & 126 &63& 63   &  63 & $\{5/9,6/9\}$ & 200    \\
I (2014.12) & 109 & 63& 54.0 & 49  & 7/9  & 200    \\
I (2015.9) & 121 & 63 & 60.5 & 58  & 8/9  & 400    \\
\hline
II (2011.1) & 33 & 33 & 33 & 33    & $\{8/15,5/9,6/9\} $ & 33    \\
\hline
\end{tabular}
\end{table*}

 Subjects answered the questions individually using their respective private
 signals and information about the previous subjects' choices.
 This information,
 called social information, was given as the summary statistics
 of the previous subjects.
 If the subject answers question $q,i$
 after $t-1$ subjects, the subject receives a
 private signal $S(q,i,t)$ and
 social information $\{C_{A}(q,i,t-1),C_{B}(q,i,t-1)\}$ from the previous
 $t-1$ subjects.
 Let $X(q,i,s)\in \{A,B\}$ be the $s$-th subject's choice;
 the social information $C_{x}(q,i,t-1),x\in \{A,B\}$
is written as
\[
C_{x}(q,i,t-1)=\sum_{s=1}^{t-1}\delta_{X(q,i,s),x},
\]
where $C_{A}(q,i,t-1)+C_{B}(q,i,t-1)=t-1$ holds.

\begin{figure}[htbp]
\begin{tabular}{c}
\includegraphics[width=11cm]{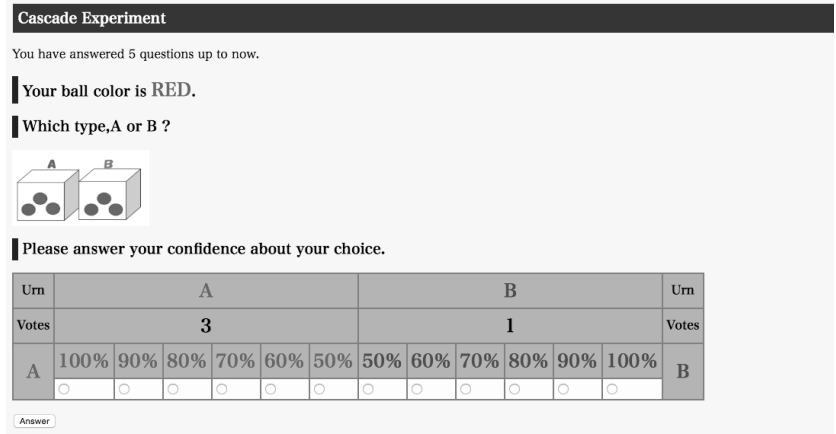}
\end{tabular}
\caption{\label{fig2}
Snapshot of the screen for
$q=6/9=2/3$ in EXP-I.
The private signal is shown on the second line.
The summary statistics $\{C_{A}(t),C_{B}(t)\}$ appear in the
second row in the box.}
\end{figure}	

  Figure \ref{fig2} illustrates the experience of 
  subjects in EXP-I more concretely. The second line shows
 the subject's private signal.
  The figure below the question shows the type of question, $q$.
 Before the experiment,
 the experimenter described the ball configuration in urns A and
 B and explained how the signal is related to the likelihood for each urn.
 The subjects can recall the question by looking at the figure.
 In the second row of the box, the social information is provided.
 In the screenshot shown in the figure, four subjects have already
 answered the question.
 Three of them have chosen urn A, and one has chosen urn B.
 The subject chooses urn A or urn B using the radio buttons in the last
 row of the box. They were asked to choose by stating how confident
 they are about their answer, that is, to choose 100\% if they were certain
 about their choice and to choose 50\% if they were not at all confident about 
 their choice. The reward for the correct choice does
 not depend on the confidence level. Irrespective of the degree of
 confidence, subjects receive a positive return for the correct choice.
 After they chose an option and put answer button, we let them know
 the correct choice in the next screen. 
 In EXP-II, the subjects were asked to choose urn A or urn B, and they
 were not asked to state their degree of confidence.
 In addition, we did not let them know the correct option.
 We only told them their total reward.
 For more details about the experimental procedure, please refer to
 the appendices.

 Hereafter, instead of A and B, we use 1 and 0 to describe the
 correct and incorrect choices and private signal as in the previous section.
 We use the same notation
 for them, as follows:
 $S(q,i,t)\in \{0,1\}$ and $X(q,i,t)\in \{0,1\}$.
 For the social information, we define $\{C_{1}(q,i,t),C_{0}(q,i,t)\}$
 as $C_{1}(q,i,t)\equiv C_{U(q,i)}(q,i,t)$ and 
 $C_{0}(q,i,t)\equiv t-C_{1}(q,i,t)$. 
 Further, $C_{1}(q,i,t)$ shows the number of correct choices
 up to the $t$-th subject for question $q\in Q,i\in \{1,\cdots,I\}$.
 In EXP-I, the length of $\{X(q,i,t)\}$ and $\{S(q,i,t)\}$ depends
 on $i\in I$ for $q=7/9$ and $8/9$, and one should write its
 dependence on $i$
 explicitly as $T(q,i)$.
 For simplicity, we use $T$ whenever it will not cause confusion.
 For example, we denote the percentage  up to the
$t$-th subject for question $q,i$ as $Z(q,i,t)$:
\[
Z(q,i,t)=\frac{1}{t}\sum_{s=1}^{t}X(q,i,s).
\]
We write the final value $Z(q,i,T(q,i))$ as $Z(q,i,T)$.

\section{\label{sec:da}Data Analysis}
In this section, we show the results of the analysis
of the experimental data. We describe how the social
information and private signal
 affect subjects' decisions.

\subsection{Distribution of $Z(q,i,T)$}
We study the relationship between the precision of the signal $q$
and $Z(q,i,T_{min})$.
As we are interested in the dependence on
the initial value of  $X(q,i,1)$, we divide the samples according
to the value of
$X(q,i,1)=x$. We denote the sample number and the
average value of $Z(q,i,t)$ for
each case $X(q,i,1)=x$ as $I(q|x)$ and $Z_{avg}(q,t|x)$, respectively.
\begin{eqnarray}
I(q|x)&=&\sum_{i=1}^{I} \delta_{X(q,i,1),x}, \nonumber \\
Z_{avg}(q,t|x)&=&\frac{\sum_{i=1}^{I}Z(q,i,t)\delta_{X(q,i,1),x}}{I(q|x)}.
\end{eqnarray}
The unconditional average value of $Z_{avg}(q,t|x)$ is then given as
\[
Z_{avg}(q,t)=q\cdot Z_{avg}(q|1)+(1-q)\cdot Z_{avg}(q|1).
\]
$Z_{avg}(q,t)$ corresponds to $P_{2}(t)$ in the simple model, and
 the deviation of $Z_{avg}(q,t)$ from $q$ is a measure
 of the collective intelligence.


\begin{figure}[htbp]
\begin{tabular}{cc}
\includegraphics[width=8.5cm]{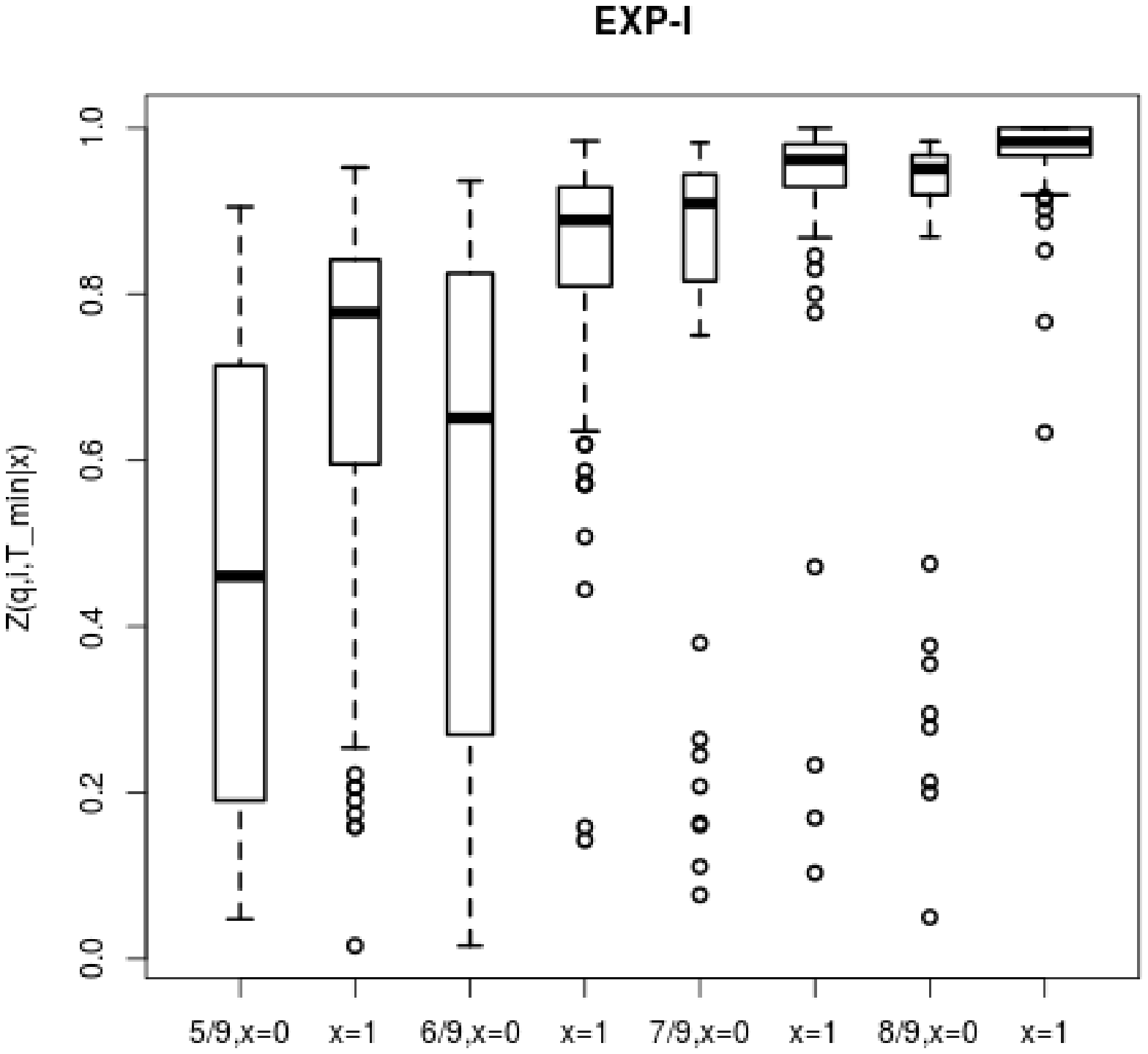} &
\includegraphics[width=8.5cm]{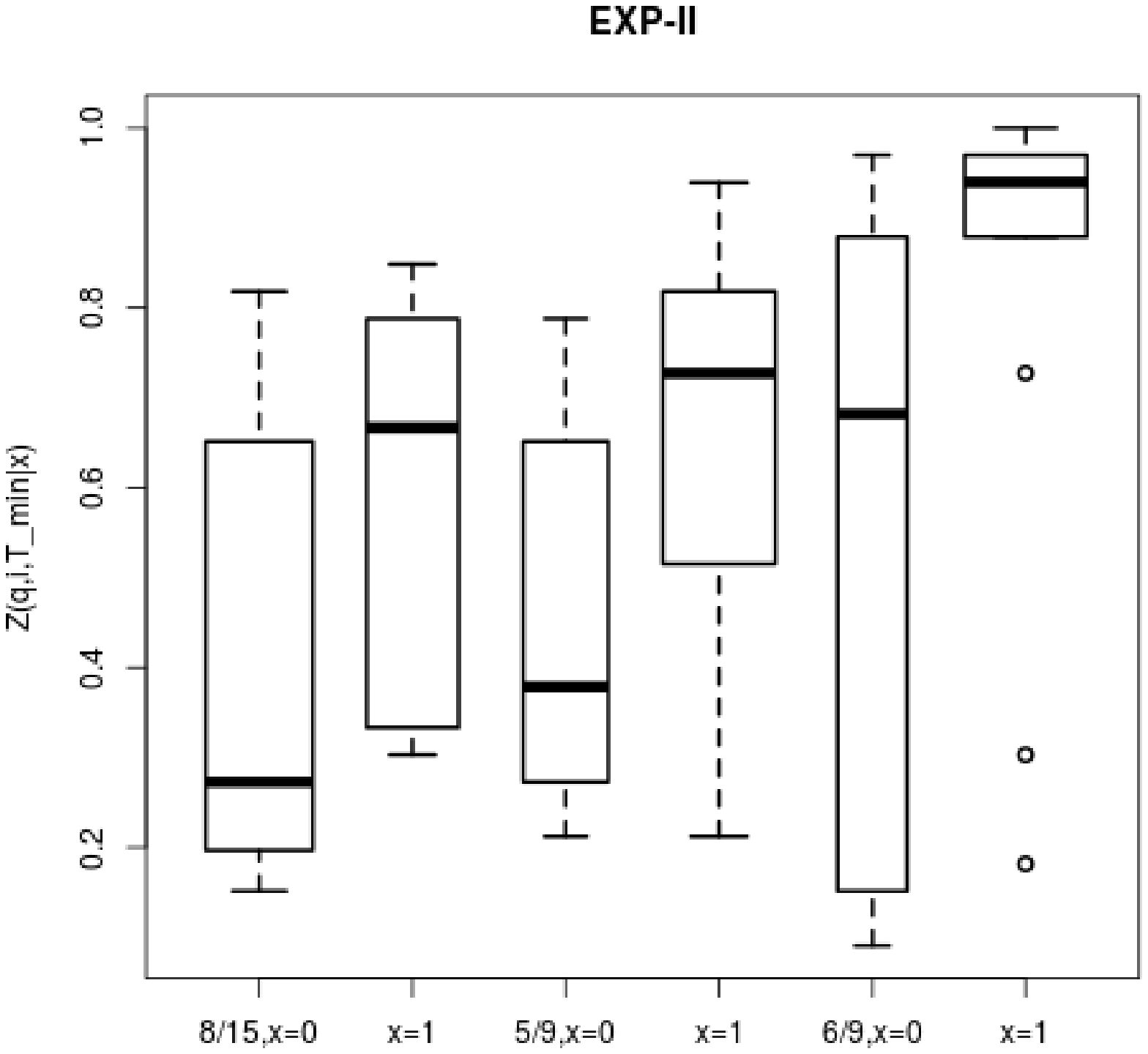}
\end{tabular}
\caption{\label{fig3}
Boxplot of $Z(q,i,T_{min})$ in EXP-I (left) and EXP-II (right).
}
\end{figure}

Figure \ref{fig3} shows boxplots of $Z(q,i,T_{min})$ for the samples with
$X(q,i,1)=x\in \{0,1\}$.
From left to right, $q$ increases. When $q$ is small, $Z_{avg}(q,T_{min}|x)$
 is small. The distribution of $Z(q,i,T_{min})$ also depends
on the initial  value
 $X(q,i,1)=x$. For $q=8/9$ in EXP-I, all $Z(q,i,T_{min})$ are 
 larger than one-half
 for $x=1$. This suggests that $Z(q,i,t)$ converges to almost 1 as
 $t$ increases. On the other hand, if $x=0$ for $q=8/9$,
 there are some samples with $Z(q,i,T_{min})<1/2$. We cannot judge whether
 all $Z(q,i,t)$ converge to almost 1 in the limit $t\to \infty$.
 If $x=0$ with $q\in \{5/9,6/9\}$,
 the distribution of $Z(q,i,T_{min})$ is wide, 
 suggesting the existence of  multiple fixed points where $Z(q,i,t)$
 converges.

\begin{figure}[htbp]
\begin{center}
\includegraphics[width=9cm]{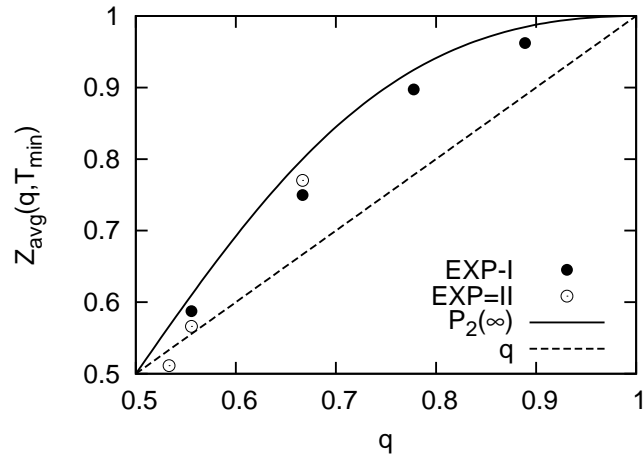}
\caption{\label{fig4} Plot of $Z_{avg}(q,T_{min})$ and $P_{2}(\infty)$ vs. $q$.
$P_{2}(\infty)$ is given by Eq. (\ref{eq:P2}).
}
\end{center}
\end{figure}	

Figure \ref{fig4} plots $Z_{avg}(q,T_{min})$ and $P_{2}(\infty)$ in
Eq. (\ref{eq:P2}) as a function of $q$. One can clearly see the collective
 intelligence effect, as $Z_{avg}(q,T_{min})-q$ is positive in almost all cases.
 For $q=8/15$ in EXP-II, the number of samples is limited and the difference
 is small, so there is no significant difference.
  One also sees that $P_{2}(\infty)$ in 
 Eq. (\ref{eq:P2}) describes $Z_{avg}(q,T_{min})$ relatively well. 
 However, it does not
 mean that the experiment should be described by the
 simple model. As we shall see below, the system shows a  phase
 transition, and the simple model is essentially wrong.

\subsection{Strength of social influence and private signal}
To measure how strongly the social information
and private signal affected subjects' decision making, we compare
the correlation coefficients between them and the subjects' decisions.
We estimate the correlation coefficients as
\begin{eqnarray}
\mbox{Cor}(S(t),X(t))&\equiv&\frac{\overline{X(t)S(t)}
-\overline{X(t)}\cdot \overline{S(t)}}
{\sqrt{\mbox{V}(X(t))\mbox{V}(S(t))}}, \nonumber \\
\mbox{Cor}(Z(t-1),X(t))&\equiv &\frac{\overline{X(t)Z(t-1)}
-\overline{X(t)}\cdot  \overline{Z(t)}}
{\sqrt{\mbox{V}(X(t))\mbox{V}(Z(t))}}, \nonumber  \\
\overline{A(t)}&\equiv & \frac{1}{I}\sum_{i=1}^{I}A(q,i,t), \nonumber \\
\mbox{V}(A(t))&\equiv &\overline{A^{2}(t)}-\overline{A(t)}^{2}. \nonumber
\end{eqnarray}
Here, we also define the average value $\overline{A}$
and variance $\mbox{V}(A)$ of quantity $A$.

\begin{figure}[htbp]
\begin{tabular}{cc}
\includegraphics[width=8.5cm]{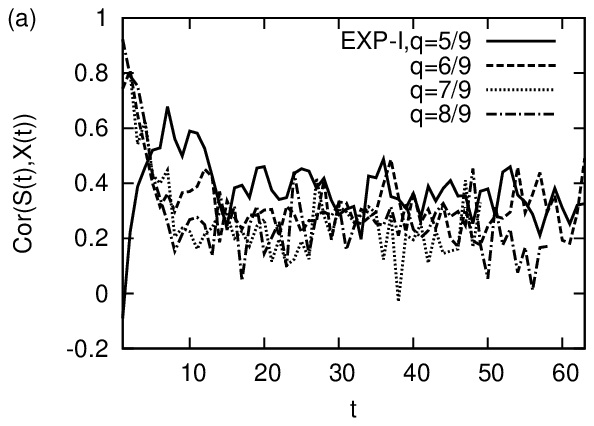} &
\includegraphics[width=8.5cm]{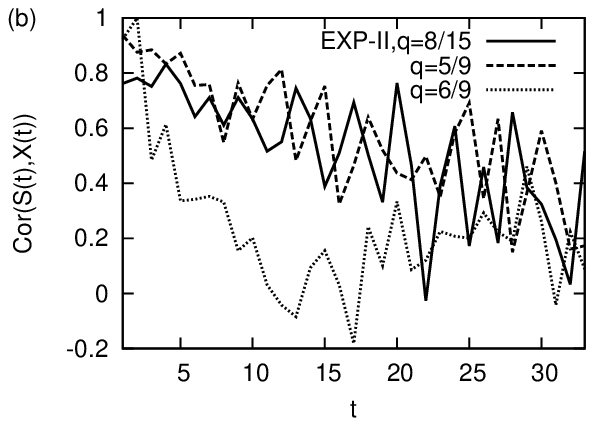} \\
\includegraphics[width=8.5cm]{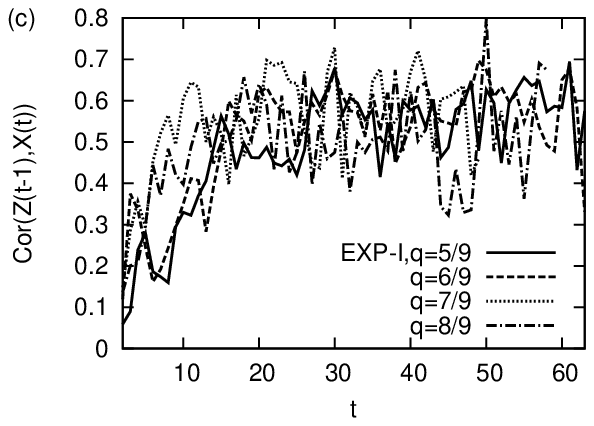} &
\includegraphics[width=8.5cm]{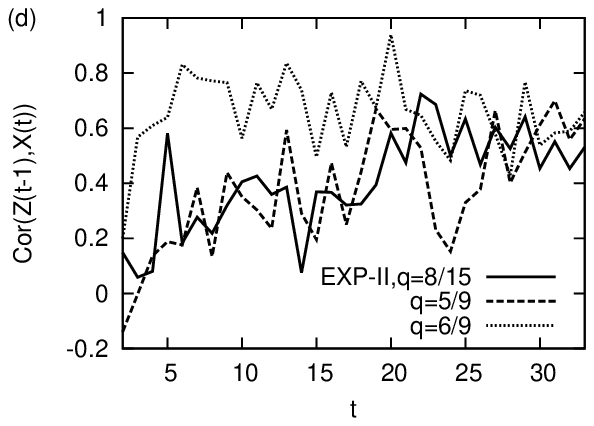}
\end{tabular}
\caption{\label{fig5}
Correlation coefficients Cor$(S(t),X(t))$
and Cor$(Z(t-1),X(t))$ vs. $t$  in (a), (c) EXP-I 
and (b), (d) EXP-II.}
\end{figure}

Figure \ref{fig5} shows plots of the correlation coefficients versus $t$.
Overall, Cor$(S(t),X(t))$ decreases  and Cor$(Z(t-1),X(t))$
 increases with increasing $t$.
In EXP-I, for $q=5/9$, Cor$(S(t),X(t))$
starts at very small values (Figure \ref{fig5}a).
We think that subjects were confused
at small $q$, and they could not trust their private signals at small $t$.
However, Cor$(S(t),X(t))$ rapidly increases and
behaves similarly to the other coefficients.
At around $t=15$, the correlation coefficients fluctuate around certain
values. The results suggest that the system becomes
stationary for $t\ge 15$. Cor$(S(t),X(t))$ and Cor$(Z(t-1),X(t))$
fluctuate around 0.3 and 0.6, respectively.
This indicates that the social influence is stronger
than the private signal.

\subsection{Response functions $f(z,s)$}
We study how subjects' decisions are affected by the social information and
private signal. We study the probabilities that $X(t+1)$ takes 1
under the condition that $Z(t)=z$ and $S(t+1)=s$. We denote them as
\[
f(z,s)\equiv \mbox{Pr}(X(t+1)=1|Z(t)=z,S(t+1)=s).
\]
By symmetry under the transformations
$S\leftrightarrow 1-S$, $X\leftrightarrow 1-X$, and $Z\leftrightarrow 1-Z$,
$f(z,s)$ has the $Z_{2}$ symmetry
\[
1-f(1-z,0)=f(z,1).
\]
In the estimation of $f(z,s)$ using experimental data
$\{S(q,i,t),X(q,i,t)\}$, we exploit
the symmetry. If $S(q,i,t)=0$, we
 replace $(S(q,i,t)=0,Z(q,i,t-1),X(q,i,t))$
 with $(1-S(q,i,t),1-Z(q,i,t-1),1-X(q,i,t))$ and estimate
$f(z,1)$.
Then $f(z,0)$ is given as $f(z,0)=1-f(1-z,1)$.
In addition, as we are interested in the static behavior of $f(z,s)$,
and Cor$(S(t),X(t))$ and Cor$(Z(t-1),X(t))$ reach their stationary
 values at $t=15$, we use data $\{S(q,i,t),X(q,i,t)\}$ for $t\ge 16$.

\begin{figure}[htbp]
\begin{tabular}{c}
\includegraphics[width=12cm]{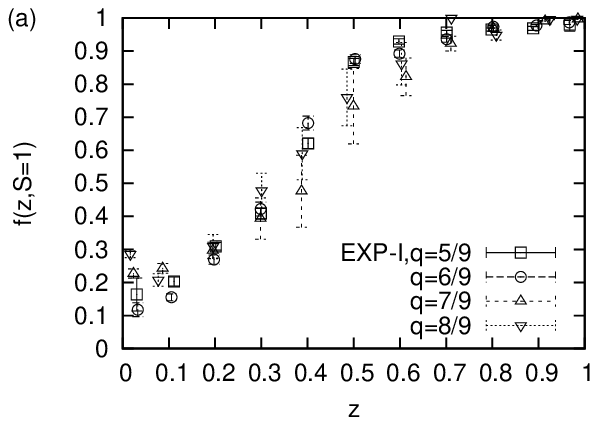} \\
\includegraphics[width=12cm]{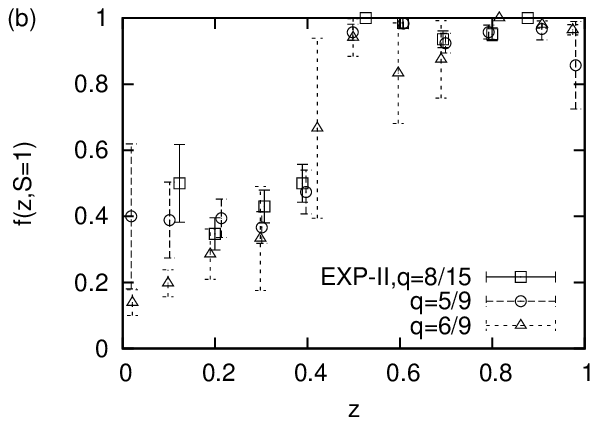}
\end{tabular}
\caption{\label{fig6}
Response functions $f(z,1)$ for $q\in Q$ in (a) EXP-I and (b) EXP-II.
$f(z,1)$ shows the probability that a subject chooses the correct urn
when $z$ percent of the previous subjects chose it and
the private signal is correct.
}
\end{figure}

We divide the samples
$\{X(q,i,t),S(q,i,t)\},16\le t\le T,i=1,\cdots,I$
according to the value of $Z(q,i,t-1)$.
We divide them into 11 bins as
$Z(q,i,t)\le 5\%,5\%<Z(q,i,t)\le 15\%,15\%<Z(q,i,t)\le 25\%,\cdots,
95\%<Z(q,i,t)$.
We write that sample $(X(q,i,t),S(q,i,t))$ is included in bin
$j\in J=\{1,2,\cdots,11\}$ as $i\in j$ and the sample number of
bin $j$ as $N(q,j)=\sum_{i\in j}1$.
We denote the average value of $Z(q,i,t)$ in bin $j$ as
$z_{j}=\sum_{i\in j}Z(q,i,t)/N(q,j)$. After this preparation, we
estimate $f(z_{j},1)$ and its error bar $\Delta f(z_{j},1)$ as
\[
f(z_{j},1)=\frac{1}{N(q,j)}\sum_{i\in j}X(q,i,t)\,\,\, , \,\,\,
\Delta f(z_{j},1)=\sqrt{\frac{f(z_{j},1)(1-f(z_{j},1))}{N(q,j)}}.
\]

Figure \ref{fig6} shows plots of $f(z_{j},1)$ versus $z_{j}$. It is clear that
$f(z_{j},1)$ are monotonically increasing functions of $z_{j}$ in EXP-I.
For $q=5/9,6/9$, their behaviors are almost the same. For $q=7/9,8/9$,
few samples appear in the middle bins, and the error bars
are large.
In EXP-II, the sample numbers are smaller than
those in EXP-I. We can see a strong positive dependence on $z_{j}$.

\section{\label{sec:da2}Detection of phase transition}
In the previous section, we introduced a response function $f(z,s)$ that
describes the probabilistic behavior of subjects in the experiments.
For $z_{j}<z<z_{j+1},i\in \{1,\cdots,10\}$, we linearly extrapolate
$f(z,s)$ with  $f(z_{j},s)$ and $f(z_{j+1},s)$.
For $z<z_{1}\,(>z_{11})$, we adopt $f(z,s)=f(z_{1},s)\,(f(z_{11},s))$.
As the private signal takes 1 with probability $q$,
 the probability that the $t+1$-th subject chooses the correct option
 under the social influence $Z(t)=z$ is estimated as
\begin{equation}
f(z)\equiv \mbox{Pr}(X(t+1)=1|Z(t)=z)
=q\cdot f(z,1)+(1-q)\cdot f(z,0) \label{eq:p_model}.
\end{equation}
We denote the averaged response function as $f(z)$. Then the
voting process $\{X(t)\},t=1,2,\cdots$
becomes a nonlinear P\'{o}lya urn process.
In this section, we study the model and verify
the possibility of a phase transition.

\subsection{Number of stable fixed points}

\begin{figure}[htbp]
\begin{tabular}{c}
\includegraphics[width=12cm]{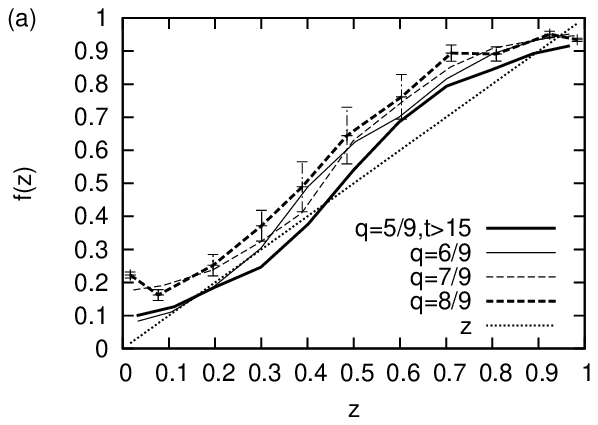} \\
\includegraphics[width=12cm]{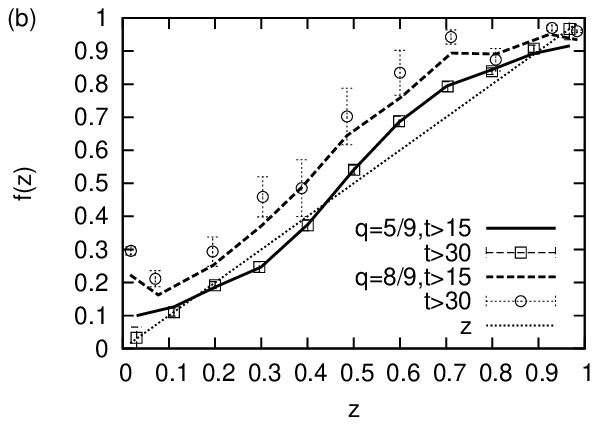}
\end{tabular}
\caption{\label{fig7}
Plot of $f(z)\equiv
\mbox{Pr}(X(t+1)=1|z(t)=z)$ in EXP-I. (a) $t> 15, q=5/9$ (thick solid line),
6/9 (thin solid line), 7/9 (thin broken line), and 8/9 (thick broken line).
(b) $q=5/9,t> 15$ (thick solid line), $q=8/9$ (thick broken line).
$q\in \{5/9,8/9\},t> 30$ with symbols.
We plot the standard error $\Delta f(z)$ for $q=8/15$ in (a) and
$t> 30$ in (b).
}
\end{figure}	

 We estimate $f(z)$ using the experimental data for EXP-I.
 We plot the results in Figure \ref{fig7}.
 For $q=5/9$ (thick solid line in Figure \ref{fig7}a), $f(z)$ crosses
 the diagonal at three points.
 The left and right fixed points are stable, and the middle one is unstable.
 Further, $z(t)$ converges to the two stable fixed points with positive probability,
 and the order parameter $c$ is positive, $c>0$. For $q=6/9$
 (thin solid line in Figure \ref{fig7}a),
 $f(z)$ touches the diagonal.
 Considering the standard error of $f(z)$,
 it is difficult to judge whether it is a touchpoint.
 However, it strongly suggests that there is another stable fixed point
 or touchpoint in addition to the right stable fixed point.
 For $q=7/9$ (thin broken line in Figure \ref{fig7}a), $f(z)$ seems to have only one stable
 fixed point. However, the departure from the diagonal is small, and
 it is difficult to judge whether there is only one stable fixed point or
 there are two stable fixed points. For $q=8/9$ (thick broken line),
 there is only one stable fixed point, and  $c$ is zero.

\subsection{Correlation function $C(t)$}
The order parameter $c$ of the phase transition is defined
as the limit value of $C(t)$.
$C(t)$ behaves asymptotically
with three parameters, $c,c'$ and $l>0$, as
\begin{equation}
C(t)\simeq c+c'\cdot t^{l-1}.  \label{eq:ct}
\end{equation}
If there is one stable state, $z_{+}$, $Z(t)$ converges to $z_{+}$.
 The memory of $X(1)=x$ in $p_{x}(t+1)$ disappears, and $c=0$.
$C(t)$ decreases to zero
with power-law behavior, $C(t)\propto t^{l-1}$.
The exponent $l$ is given by the slope of $f(x)$ at the stable
fixed point $z_{+}$ as $l=q'(z_{+})$.
If there are multiple stable states, $z_{-}<z_{+}$,
the probability that $z(t)$ converges to $z_{+}$ depends on $X(1)=x$.
If $c=\lim_{t\to\infty}(p_{1}(t+1)-p_{0}(t+1))$ is subtracted from $C(t)$,
 the remaining terms also obey a power law as $C(t)-c \propto
t^{l-1}$.
The exponent $l$ is given by the larger of $\{q'(z_{+}),q'(z_{-})\}$,
as the term with the larger value governs the asymptotic behavior of $C(t)-c$
\cite{Mor:2015-2}.
If we adopt $f(z)$ in Figure \ref{fig7}a, there are two stable states for
 $q=5/9$ and $q=6/9$. For $q=7/9$ and $q=8/9$, there is only one stable
 state. This suggests that a phase transition occurs depending 
 on $q$.

 We study the correlation function $C(t)$.
First, $p_{x}(t+1)\equiv \mbox{Pr}(X(t+1)=1|X(1)=x)$
and their error bars $\Delta p_{x}(t+1)$ are estimated
from the experimental data $\{X(q,i,t)\}$ as
\begin{eqnarray}
p_{x}(t+1)&=&
\frac{1}{N(q,x)}\sum_{i\in I}X(q,i,t+1)
\delta_{X(q,i,1),x},\nonumber \\
N_{x}(q)&=&\sum_{i\in I}\delta_{X(q,i,1),x},  \nonumber \\
\Delta p_{x}(t+1)&=&
\sqrt{\frac{p(x,t+1)(1-p_{x}(t+1))}{N_{x}(q)}}.
\nonumber
\end{eqnarray}
$C(t)$ is then estimated as
\[
C(t)=p_{1}(t+1)-p_{0}(t+1).
\]
 The standard error of $C(t)$ is given by
\begin{equation}
\Delta C(t)=\sqrt{\Delta p_{1}(t+1)^{2}+\Delta p_{0}(t+1)^{2}}.
\label{ct_error}
\end{equation}

\begin{figure}[htbp]
\begin{tabular}{cc}
\includegraphics[width=8cm]{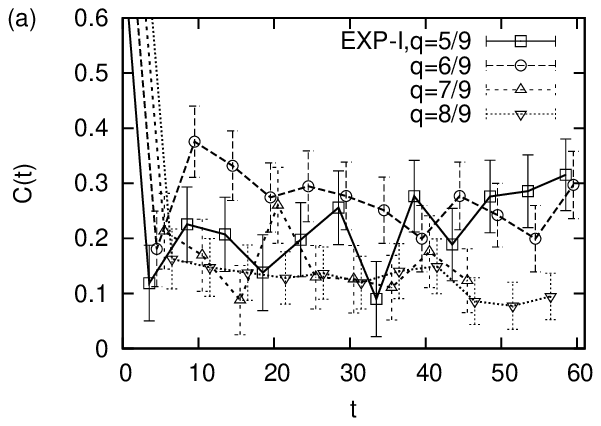} &
\includegraphics[width=8cm]{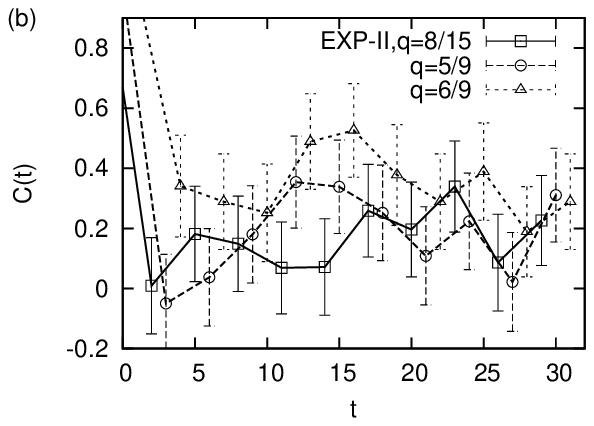}
\end{tabular}
\caption{\label{fig8}
$C(t)$ vs. $t$ in (a) EXP-I and (b) EXP-II.
Error bars are estimated using Eq. (\ref{ct_error}).
To see the behavior of $C(t)$ clearly, we plot only $C(t)$ for
$\Delta t=5 (3)$ for EXP-I (II).
In addition, we shift the data for $q=5/9,6/9 (8/15)$ leftward
and those for $q=7/9,8/9 (6/9)$ rightward for EXP-I (II).
}
\end{figure}	

Figure \ref{fig8} shows plots of $C(t)$ for $t< T_{min}$ as a function of $t$
in EXP-I and EXP-II. In both experiments, the error bars
are large. In EXP-I, $C(t)$ fluctuates around 0.25 for $q \in \{5/9,6/9\}$.
For $q\in \{7/9,8/9\}$, $C(t)$ decreases and takes small values for large $t$.
However, it is difficult to judge whether $C(t)$ decreases to zero or fluctuates
 around some positive values. In EXP-II, in all three cases, $C(t)$ seems to
fluctuate around 0.2.

\subsection{Estimation of $C(t)$ for $t\ge T_{min}$}
As the system size $T_{min}$ in our experiments is very limited,
we adopt the P\'{o}lya urn process based  on Eq. (\ref{eq:p_model})
to simulate the system for  $t> T_{min}$.
We introduce a stochastic process $\{X(t)\},t\in \{1,2,3,\cdots,T\}$.
$X(t+1) \in \{0,1\}$  is a Bernoulli random variable, and its probabilistic rule depends on all
the previous $\{X(t')\},t'\in \{1,\cdots,t\}$ through $C_{1}(t)=\sum_{t'=1}^{t}X(t')$.
The probability that $X(t+1)$ is 1 for $C_{1}(t)=n_{1}$ is given by $f(n_{1}/t)$.
We denote the  probability function for $\sum_{t'=1}^{t}X(t')=n$ with
an initial condition $X(1)=x$ as $P(t,n|x)$.
\[
P(t,n|x)\equiv \mbox{Pr}\left(\sum_{t'=1}^{t}X(t')=n|X(1)=x\right).
\]
The  master equation for $P(t,n|x)$ is
\begin{equation}
P(t+1,n|x)=f(n-1/t) \cdot P(t,n-1|x)+(1-f(n/t))\cdot P(t,n|x).
\end{equation}
We use the experimental data from EXP-I as the initial condition
for $t=T_{min}$ (Figure \ref{fig3}).
We solve the master equation recursively and
obtain $P(t,n|x)$ for $t\le 10^{6}$.
We estimate $C(t)$ as
\[
C(t)=\sum_{n=1}^{t}P(t,n|1)\cdot f(n/t)-\sum_{n=0}^{t-1}P(t,n|0)\cdot f(n/t).
\]

\begin{figure}[htbp]
\begin{center}
\includegraphics[width=12cm]{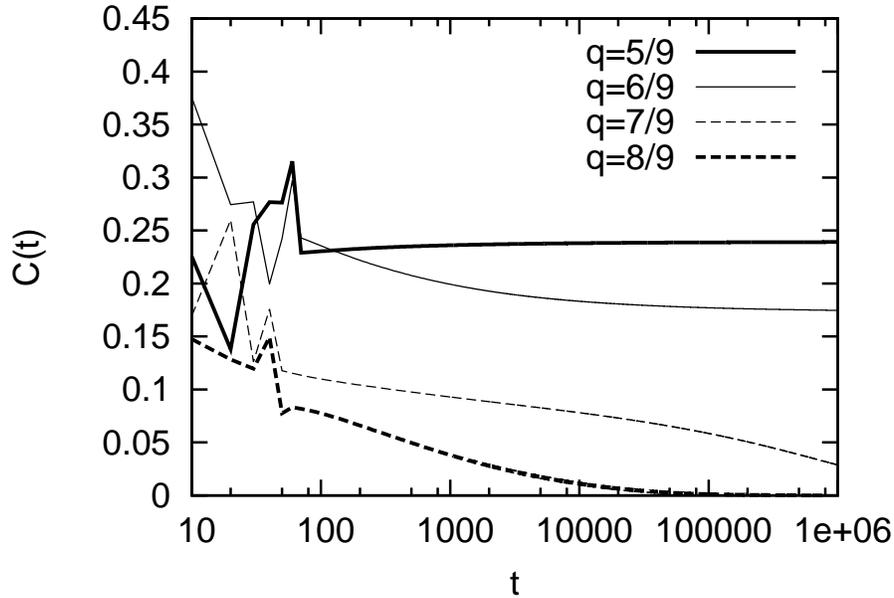} \\
\caption{\label{fig9} $C(t)$ vs. $t$ for $10^{1}\le t<10^{6}$.
For $t< T_{min}$, we plot the results in Figure \ref{fig8}a with
$\Delta t=10$.
}
\end{center}
\end{figure}

Figure \ref{fig9} shows the plots of $C(t)$ versus $t$.
For $q=5/9$, $C(t)$ converges to a finite and positive value, and $c>0$.
For $q=8/9$, $C(t)$ decays to zero very slowly. For $q=7/9$, $C(t)$
 decays more slowly, and it takes a finite value even for $t \sim 10^{6}$.
From the slope of $C(t)$ there, we can assume the limit value of
$C(t)$ is zero. For $q=6/9$, the situation is more subtle.
If $f(z)$ has a touch point, $C(t)$ decays logarithmically as
\[
C(t) \sim c+c' (\ln t)^{-1}.
\]
In this case, it is difficult to judge whether the limit value of $C(t)$
is positive or zero, as $C(t)$ decreases too slowly.
Even if it is uncertain, we can say that $c$ is positive for $q=5/9$ and
zero for $q=7,9$ and $8/9$. The system shows a phase transition.

\subsection{Estimation of $c$}
To estimate $c$, we employ the
integrated quantities of $C(t)$, which are
the integrated correlation time $\tau$ and
the second moment correlation
time $\xi$ divided by the time horizon $t$.
They are defined in terms of the moments of $C(s)$ as
\begin{eqnarray}
\tau_{t}(t)&\equiv&\tau(t)/t=m_{0}(t)/t,
\\
\xi_{t}(t)&\equiv&\xi(t)/t=
\sqrt{m_{2}(t)/m_{0}(t)},
\nonumber \\
m_{n}(t)&\equiv& \sum_{s=0}^{t-1}C(s)(s/t)^{n}.
\end{eqnarray}

\begin{figure}[htbp]
\begin{tabular}{cc}
\includegraphics[width=8cm]{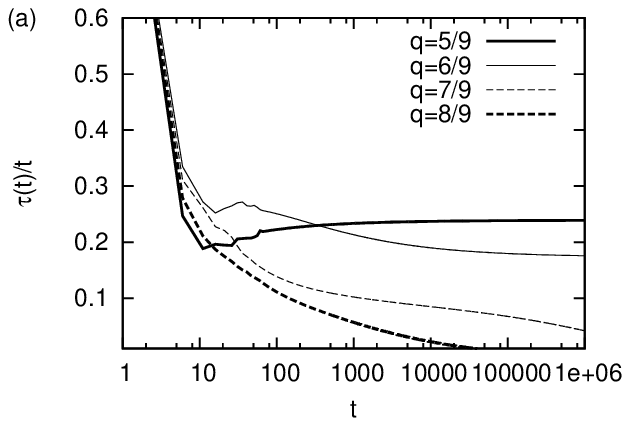} &
\includegraphics[width=8cm]{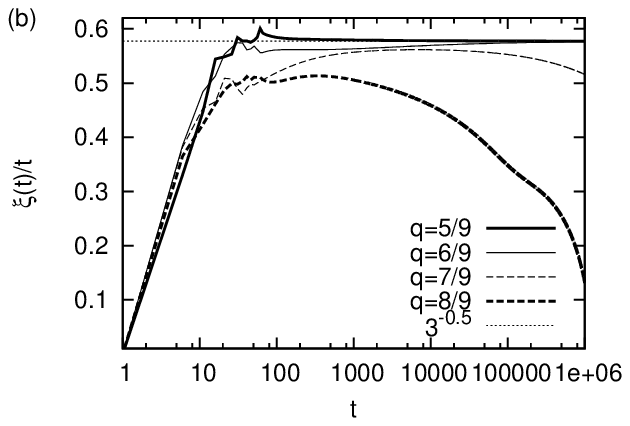} \\
\end{tabular}
\caption{\label{fig10}
Plots of (a) $\tau_{t}(t)$ and (b) $\xi_{t}(t)$ vs. $t$.
}
\end{figure}

By using the asymptotic behavior of $C(t)$ in Eq. (\ref{eq:ct}),
the limit values of $\tau_{t}(t)$ and $\xi_{t}(t)$ are found to be
\begin{eqnarray}
\lim_{t\to\infty}\tau_{t}(t)&=& \lim_{t\to \infty}c+\frac{c'}{l}t^{l-1}=c,
\label{eq:taut}
\\
\lim_{t\to\infty}\xi_{t}(t)&=&
\begin{cases}
\sqrt{\frac{l}{l+2}}\,\,\,\, , \,\,\,\, c=0, \\
\sqrt{\frac{1}{3}}\,\,\,\, , \,\,\,\,c>0.
\end{cases}
\label{eq:xit}
\end{eqnarray}
The limit value of $\tau_{t}(t)$ coincides with $c$.
With the limit value of $\xi_{t}(t)$, we can judge whether $c>0$
or $c=0$ by $\lim_{t\to\infty}\xi_{t}(t)=\sqrt{1/3}$ or
$\lim_{t\to\infty}\xi_{t}(t)<\sqrt{1/3}$.

Figure \ref{fig10} shows plots of $\tau_{t}(t)$
and $\xi_{t}(t)$ versus $t$. $\tau_{t}(t)$ increases gradually with
$t$ for $t>T_{min}$ and $q=5/9$. For sufficiently large $t$, $\tau_{t}(t)$
 for $q=5/9$ is larger than that for $q=6/9$. For $q=7/9$ and $8/9$,
 $\tau_{t}(t)$ decreases to zero monotonically, suggesting
 that $c=0$.
 $\xi_{t}(t)$ for large $t$ is smaller than $3^{-1/2}$ for
$q\in \{7/9,8/9\}$, also suggesting that $c=0$.
 For $q\in \{5/9,6/9\}$, $\xi_{t}(t)$ converges to $3^{-1/2}$
 as $t$ increases, suggesting that $c>0$.
 From these results, we conclude that
 $c$ decreases with increasing $q$ for $q\in \{5/9,6/9\}$
 and $c=0$ for $q\in \{7/9,8/9\}$.

\subsection{Plot of $P(T,n|x=0)$}
 Lastly, we show the time evolution of $P(t,n|x)$
 for the sample with $X(q,i,1)=x=0$. The boxplot of $Z(q,i,T_{min})$
 for $x=0$ in Figure \ref{fig3} shows the initial configuration
 for $P(T_{min},n|0)$.
 As there is  only one stable fixed point, $z_{+}$, for $q\in \{7/9,8/9\}$,
 $Z(q,i,t)$ should converge to $z_{+}$.
 The main interest lies in whether the samples
 with $Z(q,i,T_{min})<1/2$ for $q\in \{7/9,8/9\}$ converge to $z_{+}$.
 On the other hand, for
 $q\in \{5/9,6/9\}$, there are two stable fixed states, and 
 $P(t,n|0)$ should have two peaks.

\begin{figure}[htbp]
\begin{tabular}{cc}
\includegraphics[width=8cm]{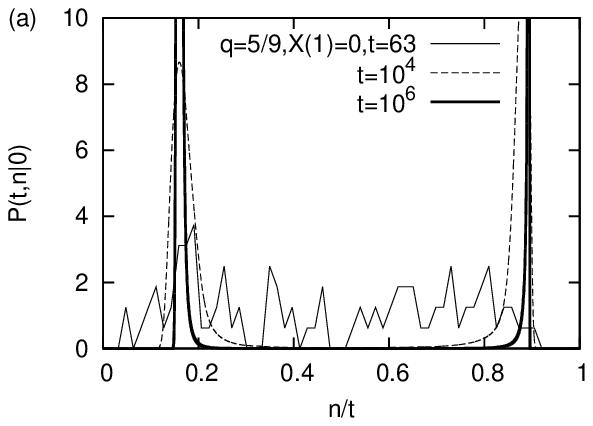} &
\includegraphics[width=8cm]{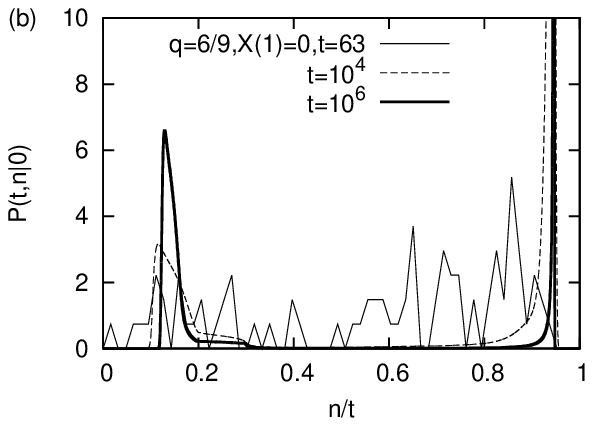} \\
\includegraphics[width=8cm]{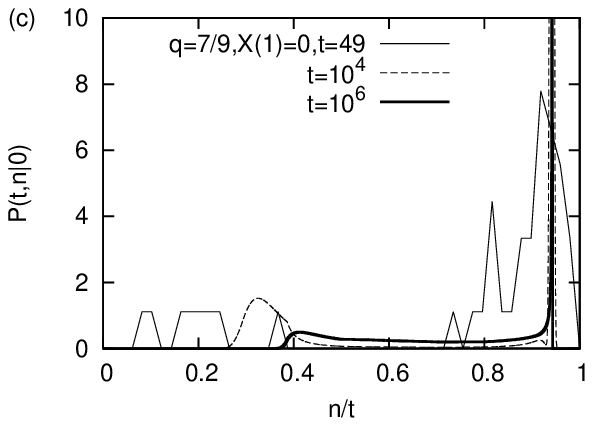} &
\includegraphics[width=8cm]{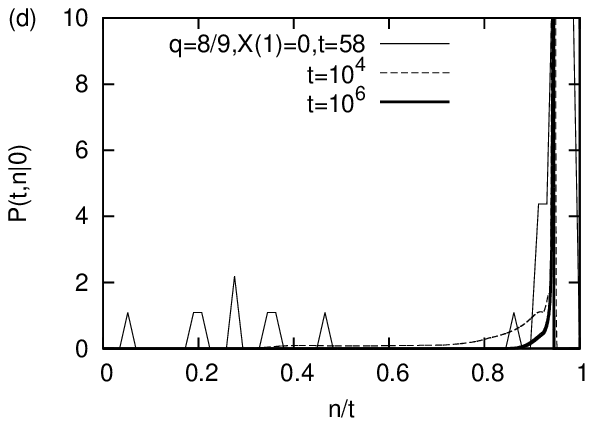}
\end{tabular}
\caption{\label{fig11}Plots of $P(t,n|x)$ with $x=X(q,i,1)=0$
 for $q\in \{5/9,6/9,7/9,8/9\}$ and $t\in \{T_{min},10^{4},10^{6}\}$.
}
\end{figure}	

Figure \ref{fig11} shows plots of $P(t,n|0)$ for
 $q\in \{5/9,6/9,7/9,8/9\}$ and $t\in \{T_{min},10^{4},10^{6}\}$.
$P(t,n|0)$ for $q\in \{5/9,6/9\}$ clearly has two peaks for $t=10^{6}$.
However, there is also a clear difference in the convergence of $P(t,n|0)$.
 For $q=5/9$. the peak at the lower stable fixed point $z_{-}$ is
 sharp for $t=10^{6}$, suggesting that the convergence is rapid.
On the other hand, for $q=6/9$, the height of the peak at the
 touchpoint is low, suggesting slow convergence.
If $f(z)$ has a touchpoint at $q_{t}$, $Z(q,i,t)$ converges to $q_{t}$ as
$|q_{t}-Z(q,i,t)|\propto (\ln t)^{-1}$ if $Z(q,i,t)$ starts below $q_{t}$.
This slow convergence is reflected in the shape of the peak at $q_{t}$.
For $q=8/9$, only one peak appears, and the sample with
 $Z(q,i,T_{min})<1/2$ converges to $z_{+}$ at $t=10^{6}$.
For $q=7/9$, as the deviation of $f(z)$ from the diagonal
 is small, the convergence to the unique stable
 fixed point $q_{+}$ is remarkably slow. Even at $t=10^{6}$, 
 a positive probability of $Z(q,i,t)<1/2$ remains. In the limit $t\to \infty$,
 the probability should disappear, and it is difficult to detect it
 experimentally.
 
\section{\label{sec:con}Summary and Comments}
 We propose a new method of detecting a
 phase transition in a nonlinear P\'{o}lya urn
 in an information cascade experiment. It is based on
 the asymptotic behavior of the correlation function
 $C(t)\simeq c+c'\cdot t^{l-1}$. The limit value
 $c$ of $C(t)$ is the order parameter of the phase transition.
 The phase transition is between the phase with $c=0$, in which
there is only one stable state,
and the phase with $c>0$, in which there is more than one stable state.
 To estimate $c$ and detect the phase transition, we propose
 to use the correlation times $\tau(t)$ and $\xi(t)$ divided by $t$.
 We perform an information cascade experiment
 to verify the method. The experimental setup is the canonical one
 in which subjects guess whether the randomly chosen urn X is urn A or urn B.
 We control the precision of the private signal $q$
 by changing the configuration of colored balls in the urns.
 We successfully detected the phase transition in the system
 when $q$ changed. For large $q$, $c=0$, and there is only one stable
 state. The system is self-correcting. For small $q$, $c>0$, and
 there are multiple stable states. The probability that the
 majority's choice is incorrect is positive.

We comment on the system size in the experiment.
In this paper, we reported on two experiments, EXP-I and EXP-II,
which differ mainly in the system size $T$ and sample number $I$.
Regarding the system size $T$, as Cor($S(t),X(t))$ and Cor$(Z(t-1),X(t))$
 fluctuate around some value for $t\ge 15$, the minimum size of $T$ should
be larger than that value in order to study the stationary behavior
of the system. Furthermore, to estimate $c$
from the asymptotic behavior of $C(t)$, it is necessary to estimate
$f(z)$ precisely. For this purpose,
$Z(q,i,t)$ should take all the values
in  $[0,1]$. As $t$ increases, $Z(q,i,t)$ converges to some stable
 fixed point of
$f(z)$. We cannot gather enough data to cover all the values
$z\in [0,1]$ if $t$ becomes too large.
Instead of setting $T$ to be large, we should set $I$ to be large.
In EXP-I, we judge that there is only one stable fixed point for $q=8/9$.
The difficulty of determining phases comes from
the error bars in the estimate of $f(z)$.
As the error bars $\Delta f(z)$ are
proportional to $1/\sqrt{I}$, $I$ should be as large as possible.
 to reduce $\Delta f(z)$.
Considering the standard errors $\Delta f(z)$ in Figure \ref{fig7},
 in order to judge whether there is only one stable fixed point for $f(z)$
 for $q=8/9$ in EXP-I, $I$ should be four times that in EXP-I.
Although $I=4\times 400=1.6\times 10^{3}$ might be large for a
laboratory experiment, it is
realizable in a web-based online experiment \cite{Sal:2006,Bon:2012}.

Another future problem is to understand and derive
the response function theoretically. A theoretical
investigation using
experimental data
for an information cascade in a two-choice general knowledge quiz was
recently performed \cite{Equ:2015}. The problem in analyzing the data for
an information cascade in a general knowledge quiz is the difficulty
in controlling the private signal \cite{Mor:2012}. The information cascade
experiment with a two-choice urn is ideal from this viewpoint. The
experimenter can control the private signal freely and
study the change in the subjects' choices.
To understand the response function, it is
necessary to control the number of referenced subjects.
We believe that experiments along these lines should be performed.
The multi-choice quiz case might be an interesting experimental subject.
In that case, the corresponding
nonlinear P\'{o}lya model is similar to the Potts model \cite{His:2015}.
The problem is whether
the herding strength increases or decreases as the number of
options changes.  We believe that the accumulation of
experimental studies in these directions is important for the development
of econophysics \cite{Man:2008,Kir:2010,Hua:2015} and
sociophysics \cite{Gal:2008,Cas:2009}.

\begin{acknowledgment}
The authors thank Ai Sugimoto, Yusuke Kishi, Kota Kuwabata, Shunsuke
Yoshida, Shion Kawasaki, and Fumiaki Sano for their support of the experiments.
The authors also thank all the participants
in the experiments.
This work was supported by JSPS KAKENHI
Grant No. 25610109.
\end{acknowledgment}

\bibliographystyle{jpsj}
\bibliography{myref}

\providecommand{\noopsort}[1]{}\providecommand{\singleletter}[1]{#1}%
\begin{thebibliography}{10}

\bibitem{Wat:2007}
D.~J. Watts: J. Consumer Research {\bfseries 34} (2007) 441.

\bibitem{Sum:2009}
D.~Sumpter and S.~C. Pratt: Phil. Trans. R. Soc. {\bfseries B364} (2009) 743.

\bibitem{Gra:2014}
J.~Fernandez-Gracia, K.~Suchecki, J.~J. Ramasco, M.~S. Miguel, and V.~M.
  Egu\'{i}luz: Phys.Rev.Lett. {\bfseries 112} (2014) 158701.

\bibitem{Sal:2006}
M.~J. Salganik, P.~S. Dodds, and D.~Watts: Science {\bfseries 311} (2006) 854.

\bibitem{Bon:2012}
R.~M. Bond, C.~J. Fariss, J.~Jones, A.~Kramer, C.~Marlow, J.E.Settle, and
  J.~Fowler: Nature {\bfseries 489} (2012) 295.

\bibitem{Wan:2014}
T.~Wang and D.Wang: Big Data {\bfseries 2} (2014) 196.

\bibitem{Pol:1931}
G.P\'{o}lya: Ann. Inst. Henri Poincar\'{e} {\bfseries 1} (1931) 117.

\bibitem{His:2006}
M.~Hisakado, K.~Kitsukawa, and S.~Mori: J. Phys. A {\bfseries 39} (2006) 15365.

\bibitem{Pem:2007}
R.~Pemantle: Pobab. Surv. {\bfseries 4} (2007) 1.

\bibitem{Art:1989}
W.~B. Arthur: Econ. Jour. {\bfseries 99} (1989) 116.

\bibitem{Hil:1980}
B.~Hill, D.~Lane, and W.~Sudderth: Ann. Prob. {\bfseries 8} (1980) 214.

\bibitem{His:2011}
M.~Hisakado and S.~Mori: J. Phys. A {\bfseries 44} (2011) 275204.

\bibitem{His:2012}
M.~Hisakado and S.~Mori: J. Phys. A {\bfseries 45} (2012) 345002.

\bibitem{Mor:2015}
S.~Mori and M.~Hisakado: J.Phys.Soc.Jpn. {\bfseries 84} (2015) 054001.

\bibitem{Mor:2015-2}
S.~Mori and M.~Hisakado: Phys.Rev. E {\bfseries 92} (2015) 052112.

\bibitem{Bik:1992}
S.~Bikhchandani, D.~Hirshleifer, and I.~Welch: J. Polit. Econ. {\bfseries 100}
  (1992) 992.

\bibitem{And:1997}
L.~R. Anderson and C.~A. Holt: Am. Econ. Rev. {\bfseries 87} (1997) 847.

\bibitem{Kub:2004}
D.~K$\ddot{u}$bler and G.~Weizs$\ddot{a}$cker: Rev. Econ. Stud. {\bfseries 71}
  (2004) 425.

\bibitem{Goe:2007}
J.~Goeree, T.~R. Palfrey, B.~W. Rogers, and R.~D. McKelvey: Rev. Econ. Stud.
  {\bfseries 74} (2007) 733.

\bibitem{Wat:2002}
D.~J. Watts: Proc. Natl. Acad. Sci. (USA) {\bfseries 99} (2002) 5766.

\bibitem{Mor:2012}
S.~Mori, M.~Hisakado, and T.~Takahashi: Phys. Rev. E {\bfseries 86} (2012)
  026109.

\bibitem{Mor:2013}
S.~Mori, M.~Hisakado, and T.~Takahashi: J.Phys.Soc.Jpn. {\bfseries 82} (2013)
  0840004.

\bibitem{Mor:2015-3}
S.~Mori, M.~Hino, M.~Hisakado, and T.~Takahashi: Proceedings of ECCS'14,
  arXiv:1507.07265 (2015).

\bibitem{Equ:2015}
V.~M. Equ\'{i}luz, N.~Masuda, and J.~Fern\'{a}ndez-Gracia: PLoS One {\bfseries
  10} (2015) e0121332.

\bibitem{His:2015}
M.~Hisakado and S.~Mori: Physica A {\bfseries 417} (2015) 63.

\bibitem{Man:2008}
R.~N. Mantegna and H.~E. Stanley: {\em Introduction to Econophysics:
  Correlations and Complexity in Finance} (Cambridge University Press,
  Cambridge, 2007).

\bibitem{Kir:2010}
A.~Kirman: {\em Complex Economics: Individual and Collective Rationality}
  (Routledge, 2010).

\bibitem{Hua:2015}
J.-P. Huang: {\em Experimental Econophysics} (Springer, 2015).

\bibitem{Gal:2008}
S.~Galam: Int. J. Mod. Phys. C {\bfseries 19} (2008) 409.

\bibitem{Cas:2009}
C.~Castellano, S.~Fortunato, and V.~Loreto: Rev.Mod.Phys. {\bfseries 81} (2009)
  591.

\end{thebibliography}

\appendix

\section{\label{sec:app:model}}
We denote the probability function  
for $N(t)\in \{-2,-1,0,1,2\}$ with the initial condition $X(1)=x$ as 
\[
P_{n}(t|x)\equiv \mbox{Pr}(N(t)=n|X(1)=x).
\]
$P_{0}(2n)$ is easily estimated as
\[
P_{0}(2n|x)= P_{0}(2|x)\cdot (2q(1-q))^{n}.
\]
$P_{\pm 2}(t)$ satisfies 
the following recursive relations for even $t$,
\begin{eqnarray}
P_{2}(2n+2|x)&=& P_{2}(2n|x)+q^{2}\cdot P_{0}(2n|x), \nonumber \\
P_{-2}(2n+2|x)&=& P_{-2}(2n|x)+(1-q)^{2}\cdot P_{0}(2n|x).  
\end{eqnarray}
$P_{\pm 1}(t)=0$ for even $t$.
For odd $t$, $P_{n}(t)$ are estimated as
\begin{eqnarray}
P_{2}(2n+1|x)&=&P_{2}(2n|x)\,\,\, ,\,\,\,  
P_{-2}(2n+1|x)= P_{-2}(2n|x), \nonumber \\
P_{1}(2n+1|x)&=&q\cdot P_{0}(2n|x)\,\,\,,\,\,\, 
P_{-1}(2n+1|x)=(1-q)\cdot P_{0}(2n|x).
\end{eqnarray}
$P_{0}(t)=0$ for odd $t$.
The initial condition for the recursive relation is
\[
P_{0}(2|x)=q\cdot \delta_{x,0}+(1-q)\cdot \delta_{x,1},
P_{2}(2|x)=q\cdot \delta_{x,1},
P_{-2}(2|x)=(1-q)\delta_{x,0}.  
\]
By solving the recursive relations with the initial condition, we have
\begin{eqnarray}
P_{2}(2n|x)&=&P_{2}(2|x)+q^{2}P_{0}(2|x)
\cdot \frac{1-(2q(1-q))^{n-1}}{1-2q(1-q)},
\nonumber \\
P_{-2}(2n|x)&=&P_{-2}(2|x)+(1-q)^{2}P_{0}(2|x)\cdot 
\frac{1-(2q(1-q))^{n-1}}{1-2q(1-q)}.
\end{eqnarray}
The unconditional probability for an up cascade is
\[
P_{2}(2n)=q\cdot P_{2}(2n|1)+(1-q)\cdot P_{2}(2n|0)
=q^{2}+q^{2}\cdot 2q(1-q) \frac{1-(2q(1-q))^{n}}{1-2q(1-q)} . 
\]
In the limit $n\to \infty$, 
it converges to 
\[
P_{2}(\infty)\equiv \lim_{n\to \infty}P_{2}(2n)
= \frac{q^{2}}{q^{2}+(1-q)^{2}}.
\]

Pr$(X(2n+1)=1|X(1)=x)$ is then estimated as
\[
\mbox{Pr}(X(2n+1)=1|X(1)=x)=P_{2}(2n|x)+q\cdot P_{0}(2n|x).
\]
$C(2n)$ is then given as
\[
C(2n)=\frac{q(1-q)}{q^{2}+(1-q)^{2}}+
\frac{(1-2q)^{2}}{2(q^{2}+(1-q)^{2})}(\sqrt{(2q(1-q))})^{2n}.   
\]
For $t=2n+1$, we can show that  $C(2n+1)=C(2n)$.

\section{\label{sec:appa}Additional information about EXP-I}
 We explain EXP-I in detail.
 We performed the experiment in 2013, 2014, and 2015.
 We recruited 126, 109, and 121 subjects in 2013, 2014, and 2015, respectively.

 In 2013, the duration of the experiment was 13 days; we recruited
 126 subjects and performed the experiment for $q\in \{5/9,6/9\}$.
 Subjects had to participate in the experiment twice.
 In the first session,
 subjects answered 100 questions for $q=6/9$.
 After a 5 min interval, they participated in another
 cascade experiment.
 In one session, a subject had to participate
 in two types of information cascade experiment.
 In the second session,
 subjects answered 100 questions for $q=5/9$.
 After a 5 min interval, they participated in another
 cascade experiment.
 The allotted time for one session was 90 min, which included
 time for an explanation of the experiment.
 The subjects received 10 yen (about 8 cents) for each correct choice.
 After they participated in two sessions for two values of $q$,
 they were given their reward.

 We performed the experiment for $q=7/9$ in 2014.
 The duration of the experiment was 13 days, and we recruited
 109 subjects.
 Thirty-nine of the subjects had participated in the experiment in 2013.
 As in the experiment in 2013, after they answered 100 questions for
 $q=7/9$, they participated in another cascade experiment.
 A problem with the web server used for the experiment occurred on
 the first day in 2014, and some
 participants could not answer all 100 questions in the
 allotted time.
 The subjects received 5 yen (about 4 cents)
 for each correct choice.

 In 2015, we performed the experiment for $q=8/9$.
 The duration of the experiment was 7 days, and we recruited
 121 subjects. In the experiment,
 the subjects answered 200 questions for
 $q=8/9$ only, and they did not participate in another experiment.
 Ten of the subjects had participated in both of the first two experiments.
 Within the allowed time of about 40 min, they could not answer
 all questions.
 The subjects received 5 yen (about 4 cents)
 for each correct choice in addition to a payment of 150 yen
 (about 1.2 dollars) for participating.

  Next, we explained the experimental procedure.
  Subjects entered a room and sat in a seat.
  There were two documents on the desk in front of the seat:
  an experimental participation consent
  document  and a brief explanation of the experiment.
  The experimenter described the experiment
  and the reward using the document. Next, the subjects
  signed the consent document and logged into the experiment's
  web site using IDs assigned by the experimenter.
  Then they started to answer the questions.
  After the experiment started, communication
  among participants was forbidden.
  A question was chosen by the server used for the experiment and displayed on
  the monitor of a 7 in. tablet (e.g., Nexus 7).
  There were no partitions in the room, and subjects could see each other.
  However, the displays on the tablets were small, and the subjects
  could not see which question the other subjects received and
  which option they chose.

\section{\label{sec:appb}Additional information about EXP-II} 
 We recruited 33 subjects for EXP-II.
 We performed the experiment in one day.
 Originally, we planned to obtain data for the experiment
 with $T=33$ and $Q=\{6/9,5/9,8/15\}$ twice within
 3 h. We prepared $I=33$ questions and the private
 signals $U(q,i,t)$ for $T$ subjects for question $q,i$.
 We let all 33 subjects enter an information science
 laboratory, and
 they participated in the experiment simultaneously.
 Subject $j=1,\cdots,33$ answered question $i=1,\cdots,I$ as
 the $t=(i+j-2)\mbox{mod}33+1$-th subject. However,
 this procedure caused a ``traffic jam,'' and
 the server used for the experiment could not serve questions smoothly.
 Within the 3 h allotted, we could gather data only for the first
 three cases, i.e., 99 questions. Subjects received 10 yen (about 8 cents) for
 each correct choice. There was a payment of 3000 yen (about
 \$25) for participating.

\section{\label{sec:appc}Asymptotic behavior of V$(z(q,t))$} 

We studied the asymptotic behavior of the variance of $Z(q,t)$ and
verified the possibility of the phase transition. In contrast to the
method based on $C(t)$, the analysis of the variance has the advantage
 that it can directly detect the existence of multiple stable
states. The drawback is the estimation of the standard errors, as
we do not know the the distribution of $Z(q,t)$.

\begin{figure}[htbp]
\begin{tabular}{cc}
\includegraphics[width=8cm]{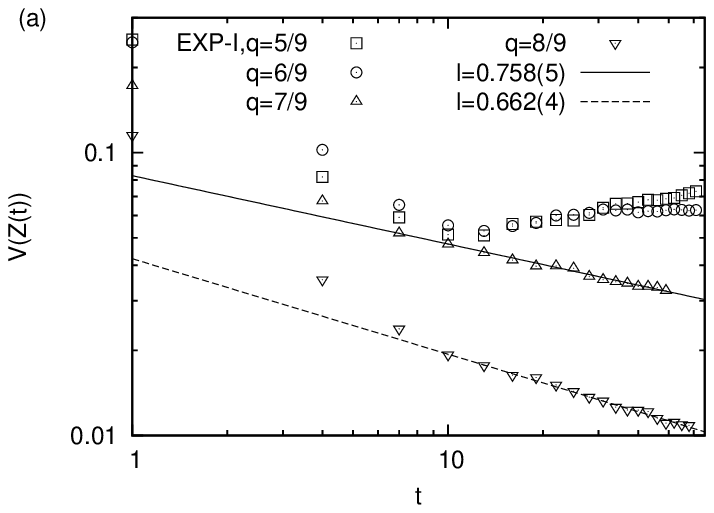} &
\includegraphics[width=8cm]{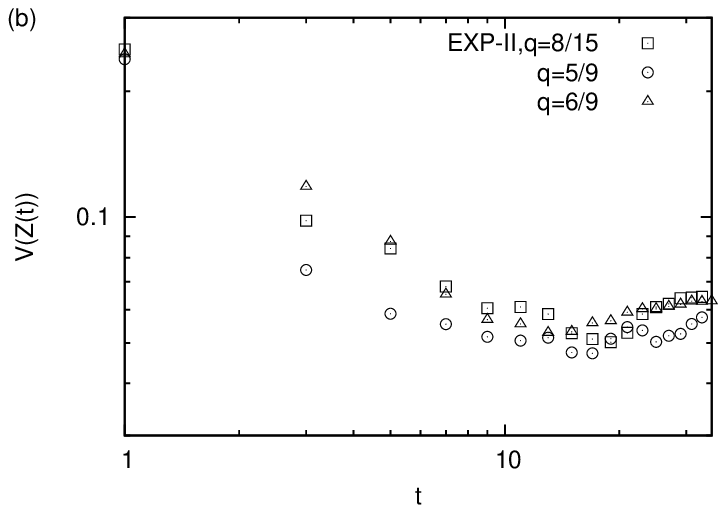}  \\
\end{tabular}
\caption{\label{fig12}
V$(Z(q,t))$ vs. $t$ in (a) EXP-I and (b) EXP-II.
Solid line in (a) shows the results fitted with
V$(Z,q,t)\propto t^{l-1}$ and $l=0.758$.}
\end{figure}

Figure \ref{fig11} shows plots  of V$(Z(q,t))$ versus $t$.
For $q\in \{5/9,6/9\}$ in EXP-I and for all cases in EXP-II,
V$(Z(q,t))$ seems to converge to some positive value for large $t$.
The result is consistent with the result that there are multiple
stable states in the system in these cases.
V$(Z(q,t))$ exhibits power-law behavior as
V$(Z(q,t))\propto t^{l-1}$ with $l=0.758(5)$ and
$0.662(4)$ for $q=7/9$ and $q=8/9$, respectively.
There is only one stable state in the system.
The asymptotic behavior of V$(Z(t))$ and
that of $C(t)$ is the same if $l>1/2$ \cite{Mor:2015-2}.

\section{\label{sec:appd}Archive of experimental data} 
In the arXiv site for this manuscript,
we uploaded the experimental data for both experiments.
The data are provided as CSV files, EXP-I.csv and EXP-II.csv.
They contain $X(q,i,t),S(q,i,t),ID(q,i,t)$, and $C(q,i,t)$
 for $q\in Q$, $i\in \{1,\cdots,|I|\}$, and $t\in \{1,\cdots,T\}$.
Here $C(q,i,t)\in \{50\%,60\%,\cdots,100\%\}$
indicates the confidence of the subject regarding the choice $X(q,i,t)$.
In EXP-II, the subject chose A or B directly instead of in terms of the confidence level,
so there are no data for the confidence.
$ID(q,i,t)$ are the identification numbers of the subjects.
 In EXP-II, $ID\in \{1,\cdots,33\}$, as there were 33 subjects.
 In EXP-I, in 2013, there were 126 subjects, and we labeled
them as $ID\in \{1,\cdots,126\}$. In 2014, there were 109 subjects, 39 of whom had participated in the first period. We used the same IDs for these 39 subjects and labeled the remaining 70 subjects as
$ID \in \{127,\cdots,196\}$.  In 2015, there were 121 subjects,
10 of whom participated in both experiments in 2013 and 2014.
We labeled the remaining 111 subjects as
$ID \in \{197,\cdots,307\}$.

The first column in the data file is $n$ in $q=n/(n+m)$, 
the second column is $i$,
the third column is $t$, the fourth column is $X(q,i,t)$, the fifth column
is $S(q,i,t)$, the sixth column is $ID(q,i,t)$, and the last column is $C(q,i,t)$.

\end{document}